# TCAD Simulation and Design Optimization of Radiation Hard n-MCz and n-Fz Si Microstrip Detector for the HL-LHC

**Balwinder Kaur[a], Shilpa Patyal[a], Nitu Saini[a],Puspita Chatterji[a], Ajay K. Srivastava[a,*]**

[a]Department of Physics, University Institute of Sciences, Chandigarh University, Gharuan-Mohali, Punjab, 140413, India.

*E-mail*: kumar.uis@cumail.in

ABSTRACT: A radiation hard Si detector is used in the new CMS tracker detector at HL-LHC. It has been observed that n-MCz and n-Fz Si as a material can be used for the Si micro strip detector. The detector design for this material should be simulated and optimized to get high CCE. In order to understand the charge collection behavior of the n-MCz/n-FzSi detector, it is required to simulate and compare the radiation damage effects in the mixed irradiated n-MCz Si and neutron irradiated n-FzSi micro strip detector equipped with metal overhang and multiple guard rings.

In this paper, we have done analysis and optimization of the radiation hard n-MCz Si/n-Fz Si strip detector design for the HL-LHC experiment in order to get high CCE.

KEYWORDS:*n-Fz/MCz* Si microstrip detector; TCAD simulation; Bulk damage; Full depletion voltage; Leakage current; CCE; Mixed irradiation.

* Corresponding author. Tel.: +91-8400622542; e-mail: kumar.uis@cumail.in

## Contents

## 1. Introduction

Radiation damage of the siliconstrip detectors in upcoming hadron collider poses a major issue for its reliable operation for the long–term of the experiment.The combined radiation damage effects in the Si strip detectors in the new Compact Muon Solenid (CMS) tracker at the High-Luminosity Large Hadron Collider must be investigated (HL-LHC) [1].Forn in p Si detectors exhibit good performance for the new CMS Tracker at HL-LHC experiments in the CERN RD50 European collaboration. [2]. The n-MCz (Magnetic Czochralski) Si material may be a good choice for the p in n Si strip detector. [3-6]. As a result, radiation damage effects in thin p in n-MCz Silicon (Si) microstrip detectors must be investigated. [3-6]. It was necessary to examine the macroscopic electrical performances of the p in n-MCz Si strip detectors for the new CMS tracker in order to design, develop, and optimise detectors for the HL-LHC investigations. The macroscopic and microscopic performance of these Si strip detectors measured using Current-Voltage (I/V), Capacitance-Voltage (C/V), Thermally Stimulated Current (TSC), Deep Level Transient spectroscopy (DLTS), Transient Current Technique (TCT) and Alibaba system SL, Barcelona, Spain set up.
Within the CERN RD50 collaboration, a few detector groups have developed two/three deep traps radiation damage models for p-type semiconductor (SFz) material. For the bulk of the SFz (Si) detector, the model showed good observations of the experimental and simulation result for the fluencies of the order of $1x10^{15}n_{eq}/cm^2$ 1Mev equivalent neutrons [7]. The n-MCz four deep trap model has been developed to simulate the radiation damage effects in mixed irradiated detectors up to $8x10^{14}n_{eq}/cm^2$1 Mev equal neutrons [7]. This work presents SRH calculation for the Full Depletion Voltage and leakage current for the n-Fz/n-MCz thick/thin silicon strip detector. In this work we have also used Hamberg Penta Trap Model [8]for proton radiation damage effect in n-Fz Silicon thin/thick strip detector and mixed irrdiated four level deep trap model in nMCz thin/thick Si strip[7] detector.
The paper has discussed in three sections; section 2 Hamburg Penta Trap Model (HPTM) [8] for Proton irradiation in n-Fz Si strip detector and four level deep trap mixed irradiated radiation damage model [7] for n-MCz silicon strip detector, Schockley Read Hall (SRH) calculations for the E(30K), V3, Ip, H220 and CiOi.Section 3 elaborates the results and discussion on the extrapolated values on the full depletion voltage and leakage current for mixed irridiated n-MCz thin/thick silicon strip detector and Proton irradiated n-Fz thin/thick Silicon strip detectror. In section 4 we did the TCAD simulation to get the electric field behaviour of n-Fz/n-MCz silicon strip detector.

## 2.Hamburg Penta Trap Model (HPTM) for Proton irradiation in n-Fz Si strip detector and four level deep trap mixed irradiated radiation damage model for n-MCz Si strip detector

In this section, we have used the Hamburg Penta Trap Model (HPTM) for proton irradiation in n-Fz strip detector and four level deep-trap mixed irradiation model for n-MCz Silicon strip detector for the radiation damage analysis in nFz , and nMCz Si strip detector for the HL-LHC.

### 2.1 Silicon strip detector model and SRH calculations

A rectangular cell of 0.0625 $cm^2$ x 300/200μm n-MCz Silicon strip detector model is used for the SRH calculations (see Figure 1) and the device and process parameters of the detector are shown in (Table 1). The effective doping concentration, $N_{eff}$, and generation leakage current , $I_{gen}$ at full depletion voltage, $V_{FD}$ can be calculated by the SRH theoretical expression [5], where symbols are having their usual meaning. Figure 1 shows n-MCz/n-Fz silicon stripdetector/PAD diodes irradiated with mixed irradiation on which a high positive voltage $V_0$ is applied on the $n^+$ side and $p^+$ is at ground. Device simulations have been performed with ATLAS TCAD device simulator [9]. The following physical models have been used for numerical simulation: these are Shockley–Read–Hall recombination, FLDMOB, CONMOB, Band gap narrowing, Trap Assisted Model (Hurks Model). Figure 2 show the proposed multiple guard rings layout for n-MCz Silicon Strip detector.

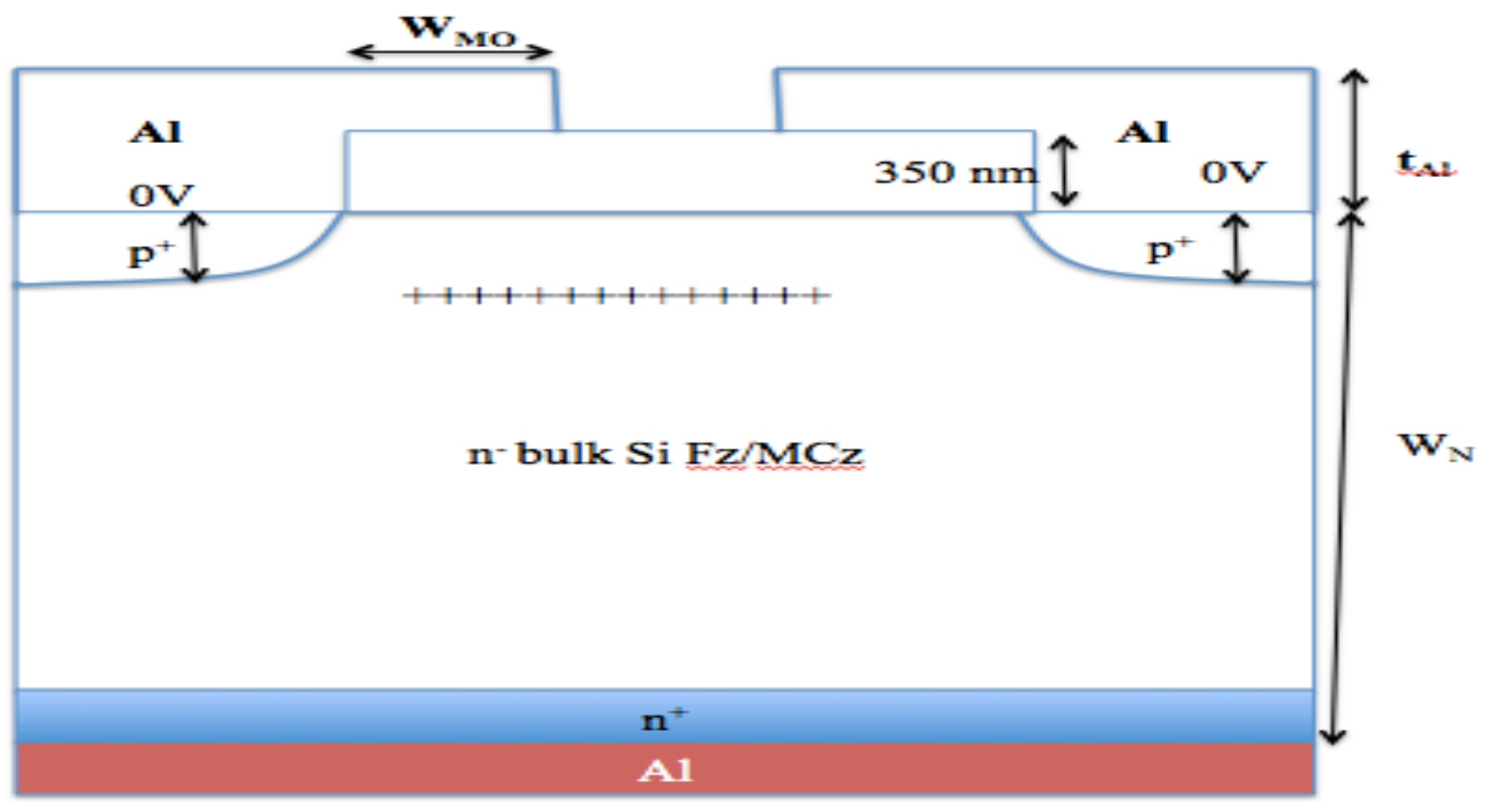

**Figure 1**. **Cross-section of the 0.0625 $cm^2$ x 200/300μm n-Fz/n-MCz Si strip detector model used in the present study for SRH/ calculations and TCAD device simulation.**

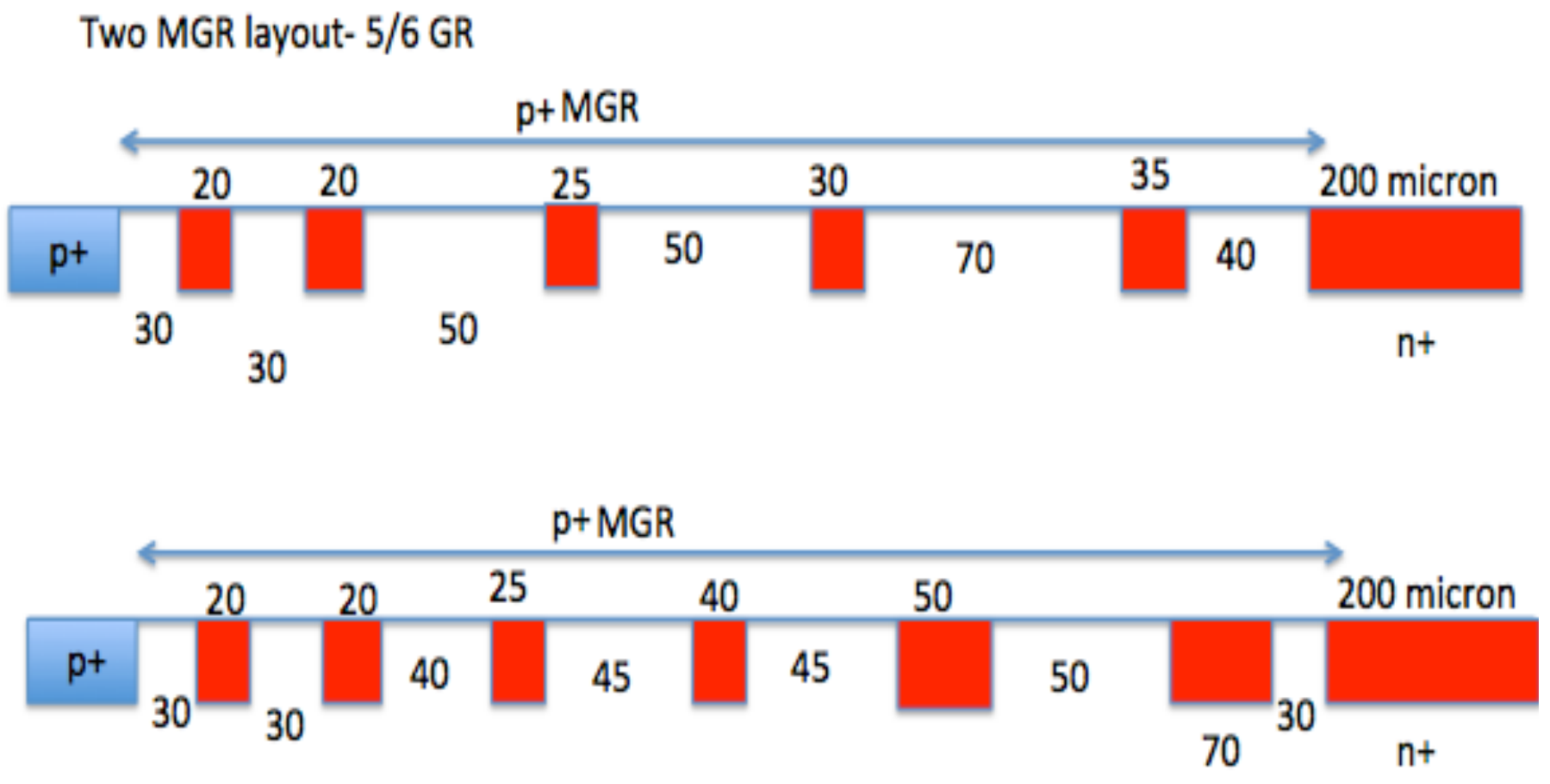

**Figure 2. Multiple Guard Rings Layout for the n-Mcz Silicon Strip Detector (Proposed)**

**Table 1**. Device and process parameters of n-MCz/n-Fz SiliconStrip detector.

| S.No. | Physical parameters | Values |
|---|---|---|
| 1. | Doping concentration ($N_D$) | 5 x$10^{12}$ $cm^{-3}$ |
| 2. | Oxide +nitride thickness ($t_{ox}$) | 0.3+.05 μm, nitride added to prevent physical damage on the interface surface |
| 3. | Junction Depth ($X_j$) | 1 μm |
| 4. | Device depth ($W_n$) | 200/300 μm |
| 5. | Fixed oxide charge ($Q_f$) | 1.5x $10^{12}$ $cm^{-2}$ |

### 2.2 SRH calculation for Full Depletion voltage andgeneration currentfor n-MCz Si Strip detector.

By using the Shockley Read Hall (SRH) recombination formula the effective doping concentration ($N_{eff}$) /full depletion voltage ($V_{FD}$) and the leakage current ($I_L$) can easily be calculated;

$$N_{eff} = N_D + \sum n_T{}^{donor} - \sum n_T{}^{acceptor} \qquad (1)$$

$$\text{Here,} \quad n_T = N_T \frac{e_{n,p}}{e_n + e_p} (2)$$

Where

$$e_p = C_p N_V \exp^{\frac{-E_t - E_V}{kT}}$$

$$e_n = C_V N_C \exp^{\frac{E_t - E_C}{kT}}$$

Here,$N_D$ is the doping concentration, $n_T$ is the steady state occupancy of defect level, $N_T$ is the defect concentration, $e_{n,p}$ is the emission rate of electron or holes, $C_{n,p}$ is the capture coefficient of electron or holes, $V_{th}$ is the thermal velocity.

After calculating the effective doping concentration, the $V_{FD}$ can be calculated by using the formula;

$$V_{FD} = \frac{{W_N}^2 N_D q}{2\epsilon_{Si}} \quad (3)$$

Here,$W_N$ is the thickness of the detector and$\epsilon_{Si}$ is the dielectric constant of silicon.

By using Shockley Read Hall recombination formula the generation leakage current ($I_L$) can be calculated;

$$I_L = qAd\left(\sum n_T^{acceptor} e_n + \sum n_T^{donor} e_p\right) = \frac{qn_iAW_N}{\tau_g} \quad (4)$$

Here, $I_L$ is generation leakage current, q is elementary charge, A is the area of the detector, d is the thickness of the detector, $n_i$ is the intrinsic carrier concentration and $\tau_g$ is the generation lifetime.

Current damage constant ($\propto$) can be calculated by using the formula:

$$\propto = \frac{A}{\emptyset_{eq} \times V} (5)$$

Here, $\propto$ is current-damage constant, $\emptyset_{eq}$ is equivalent fluence and V is the volume of the detector.

## 2.3 Hamburg Penta Trap model for protonradiation damage effect in n-Fz thin/thick Si strip detector

In this paper, we have used Hamburg Penta Trap model (HPTM) for proton radiation damage effect (See Table.2) for the radiation damage analysis in proton irradiated in n Fz Si Strip Detectorand a Four Level Deep Trap Mixed Irradiated Radiation Damage Models [7] in mixed irradiated n-type MCz Si Strip Detector.

**Table 2** Hamberg Penta Trap Model for Proton Irradiation

| Defect | Type | Energy | $g_{int}$ ($cm^{-1}$) | $\sigma_n$ ($cm^2$) | $\sigma_p$ ($cm^2$) |
|---|---|---|---|---|---|
| **E30K** | Donor | $E_C$-0.1 eV | 0.0497 | 2.300E-14 | 2.920E-16 |
| **$V_3$** | Acceptor | $E_C$-0.458 eV | 0.6447 | 2.551E-14 | 1.511E-13 |
| **$I_p$** | Donor | $E_C$-0.545 eV | 0.4335 | 4.478E-15 | 6.709E-15 |
| **H220** | Donor | $E_V$ +0.48 eV | 0.5978 | 4.166E-15 | 1.965E-16 |
| **CiOi** | Donor | $E_V$ +0.36 eV | 0.3780 | 3.230E-17 | 2.036E-14 |

## 3. Result and Discussion

In this section, we have used the SRH calculations and TCAD device simulation for the CCE comparison of the n-Fz/n-MCz thick and thin radiated Strip Detector.

### 3.1 **Full depletion voltage and leakage current obtained after the SRH Calculation for n-Fz/n-MCz thick and thin Si Strip Detector.**

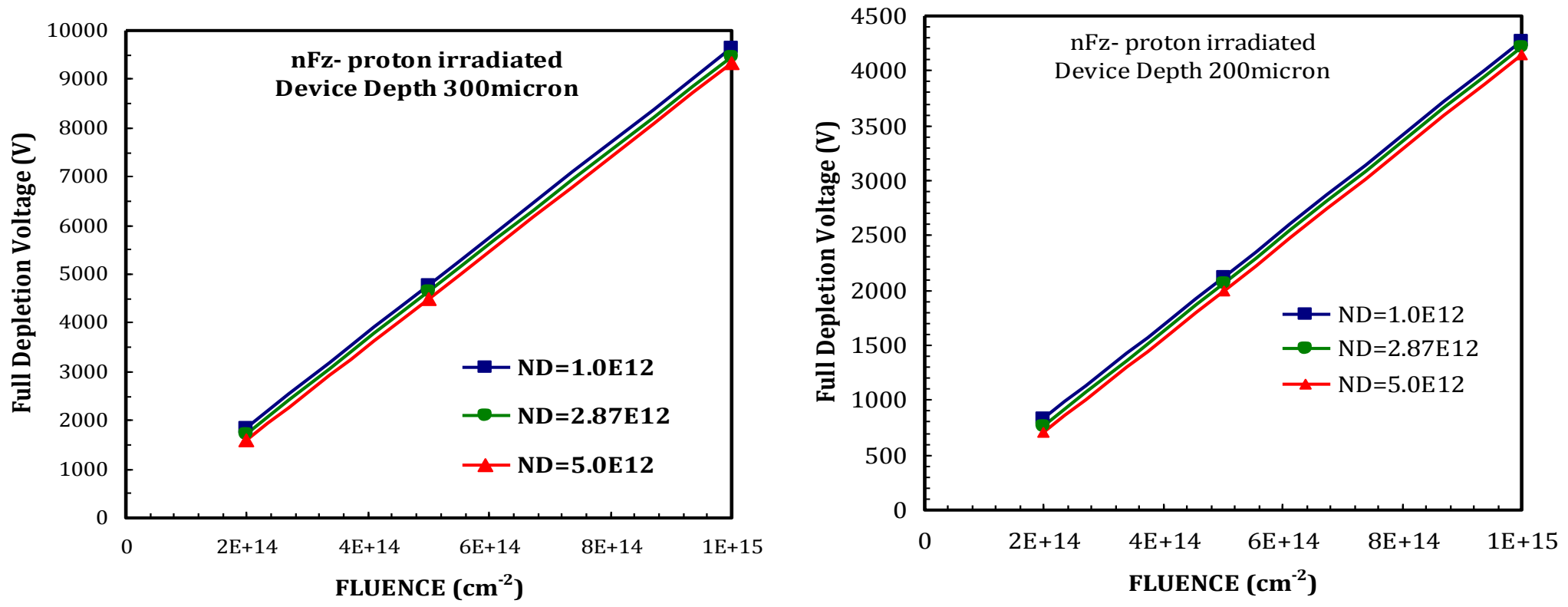

(a) (b)

**Figure 3. (a)** Full depletion voltage as a function of the Fluence in Proton Irradiated n-Fz thick Silicon (Si) Strip Detector. (b) Full depletion voltage as a function of the Fluence in Proton Irradiated n-Fz thinSilicon (Si) Strip Detector.

Hamburg Penta Trap Model (HPTM) model can reproduce the experimental data in n in p, In figure 3 a we used in 300 µm n-Fz proton irradiated Si strip detector as shown in [5], $V_{fd}$ estimated from SRH calculations [8] and decreases with increasing doping concentration Significant increase in $V_{fd}$ with the proton radiation fleunces (Φeq, n (fluence), equivalent to 1 MeV neutron) in n-Fz detector using HPTM model [5].

In figure 3 (b), it is shown that HPTM model can used to get $V_{fd}$ in 200 µm n-Fz proton irradiated Si strip detector, 50% less $V_{fd}$ have been obtained in 200 µm n-Fz proton irradiated Si strip detector than 300 µm strip detector irradiated by protons.
It is also observed from the figure significant increase in $V_{fd}$ using HPTM SRH calculations in the proton irradiated detector by 24 GeV/c protons, HPTM needs to tune the parameters to reproduce the macroscopic measurements for the good agreement in the experimental data and simulation data on n-Fz Strip detector/diodes too using silvaco TCAD.
For the $5x10^{12}$ cm$^{-3}$ doping of n-Fz bulk, less full depletion voltage (705V) obtained for the fluence of $2x10^{14}$ cm$^{-2}$than other fluencies.

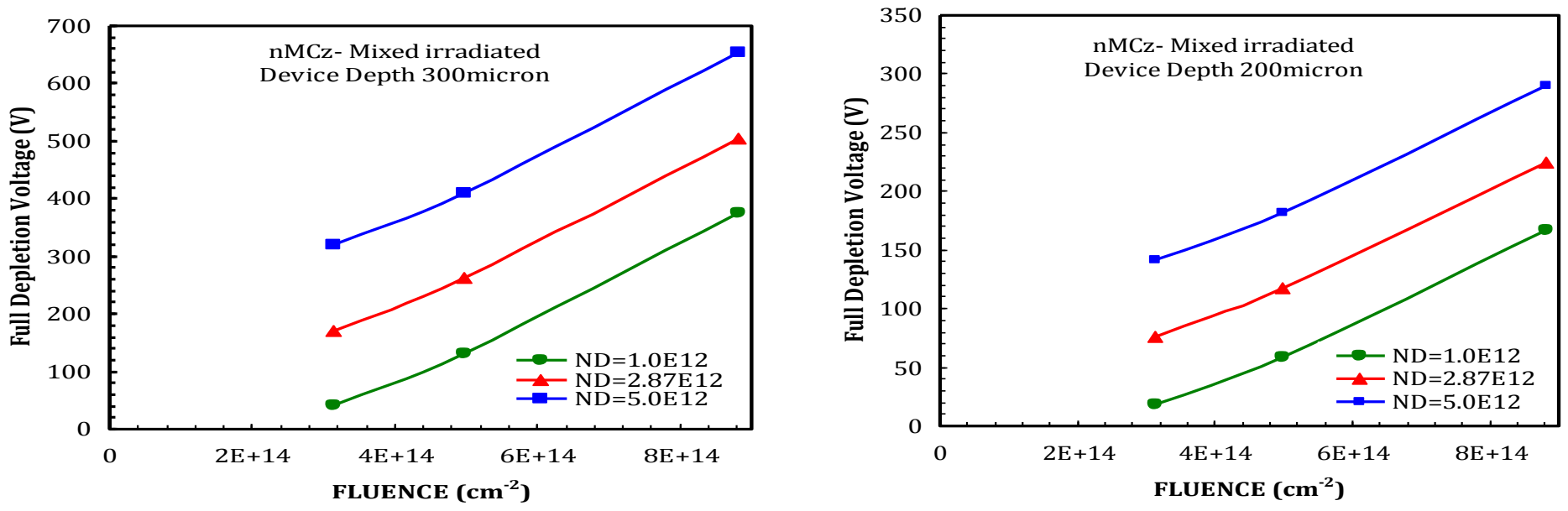

(a) (b)

**Figure 4 (a)**. Full Depletion Voltage as a function of the Fluence in Mixed Irradiated n-MCz Thick Si Strip Detector (b) Full Depletion Voltage as a function of the Fluence in Mixed Irradiated n-MCz Thin Si Strip Detector.

.

It is observed from the figure 4 (a) that $V_{fd}$ increases with the mixed irradiated fleunces for the three doping concentrations in thick n-MCz Si strip detector, less full depletion voltage obtained than other n-Fz proton radiated detectors, less $V_{fd}$ obsed due to the compensation of the deep traps in n-MCz as compared to n-Fz Si strip detector.

Full Depletion Voltage as a function of the Fluence in Mixed Irradiated n-MCz Thin Si Strip Detector is shown in figure 4 (b). It is depicted from the figure that $V_{fd}$ increases with the mixed irradiated fleunces for the three doping concentrations in thin n-MCz Si strip detector too, less full depletion voltage obtained than 300 μm Mcz Si strip detectors , less $V_{fd}$ observed in the $1x10^{12}$ $cm^{-3}$ n-MCz bulk doping, although for comparison and isolation in between strips (n+ in p detector) of the effect on the macroscopic performance and E-field distribution, have taken high doping $5x10^{12}$ $cm^{-3}$ in the n-MCz strip detector design that is giving < 300 V full depletion voltage.Thin n-MCz Si strip detector can be operated at an applied bias of 500 V.

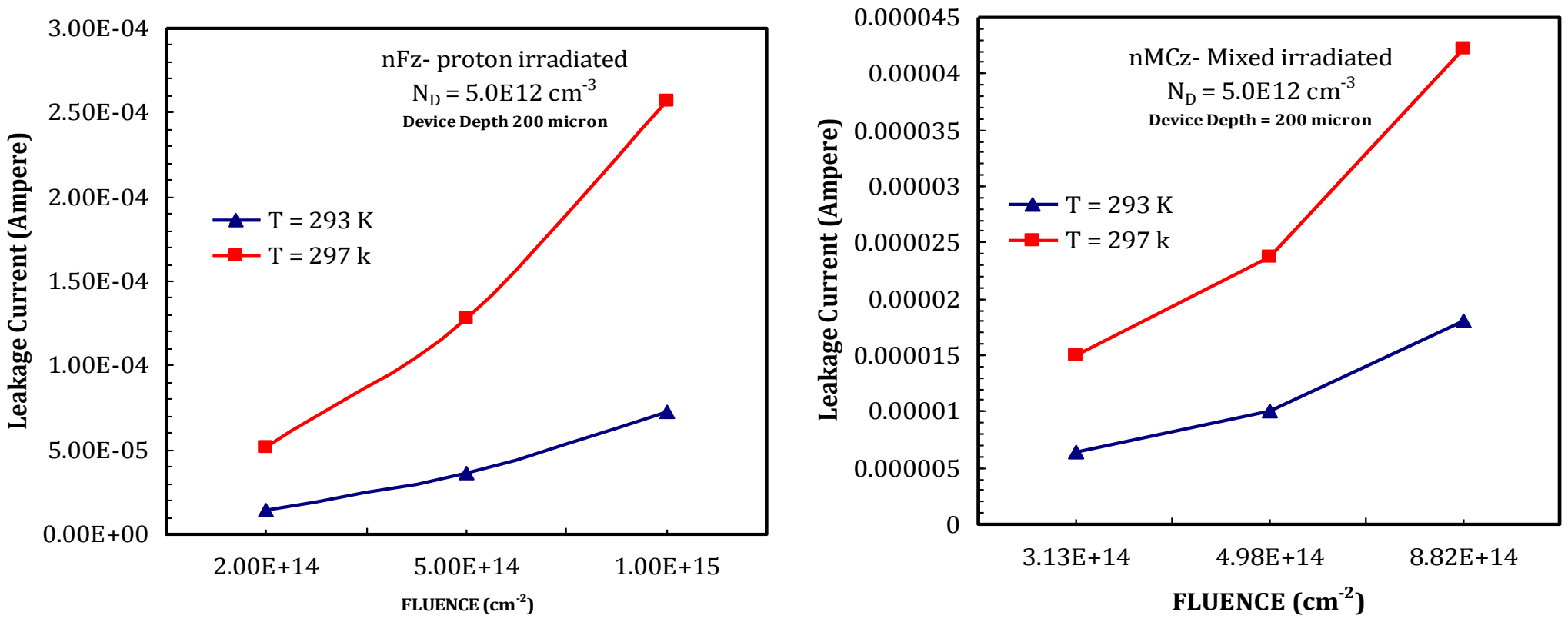

**(a)** **(b)**

**Figure 5 (a)**. Leakage Current as a function of the Fluence in Proton Irradiated Fz thin Si Strip Detector at two Temperatures (b) Leakage Current as a Function of the Fluence in Mixed Irradiated n-MCz Thin Si Strip Detector at two Temperatures.

Figure 5 (a) shows that Leakage Current as a function of the Fluence in Proton Irradiated Fz thin Si Strip Detector at two Temperatures. From the figure it is observed that the leakage current increases with fleunces at 293 (RT) , and 297 K (RT+4K) that shows the experimental results as per our work [3].

Leakage Current as a Function of the Fluence in Mixed Irradiated n-MCz Thin Si Strip Detector at two Temperatures is shown in figure 5 (b). It is evident from the figure thatLeakage current increases with fleunces at 293 (RT) , and 297 K (Rt+4K) that shows the experimental results a per[8]. It is also observed from the figure that less leakage current showed in the n-Mcz Si strip detector than n-Fz strip detector at two temperatures (292 K, 297 K).

## 4. TCAD simulation of n-MCz and n-Fz Si Microstrip Detector for the HL-LHC

**In this section, the detector model (See Figure.6) is simulated using ATLAS Silvaco and the results are shown**for the CCE comparison of the proton irradiated n-Fz and mixed irradiated n-MCz thick and thin radiated Strip Detector.

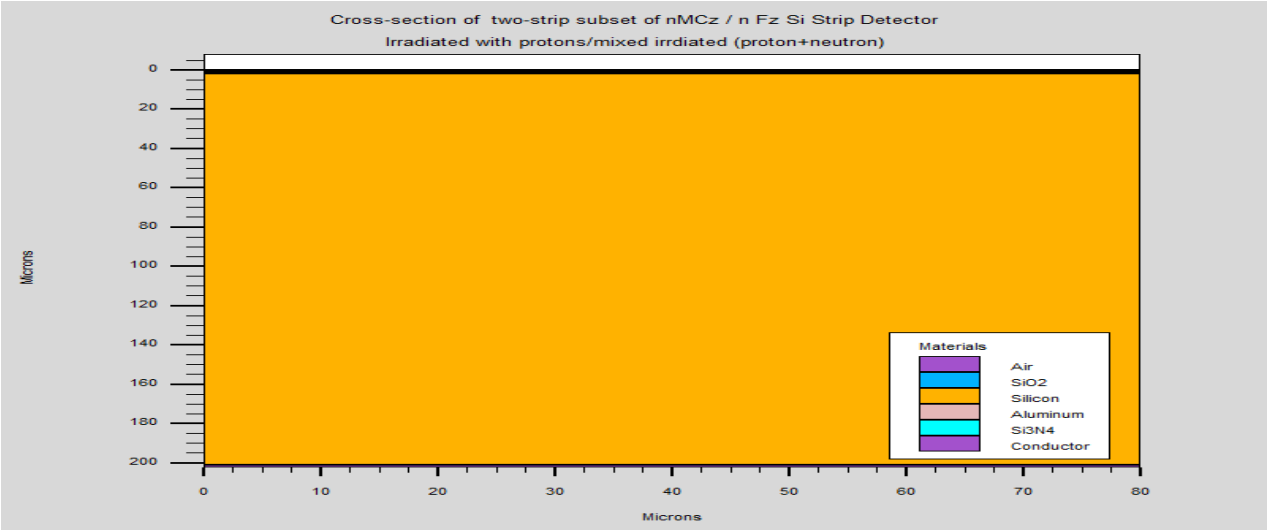

**Figure 6.**Cross-section view of two strip subset of n-MCz/n-Fz Strip Detector irradiated with protons/mixed irradiated (proton+neutrons).

## 4.1 Electric Field of Mixed Irradiated n-MCz Thin Si Strip Detector

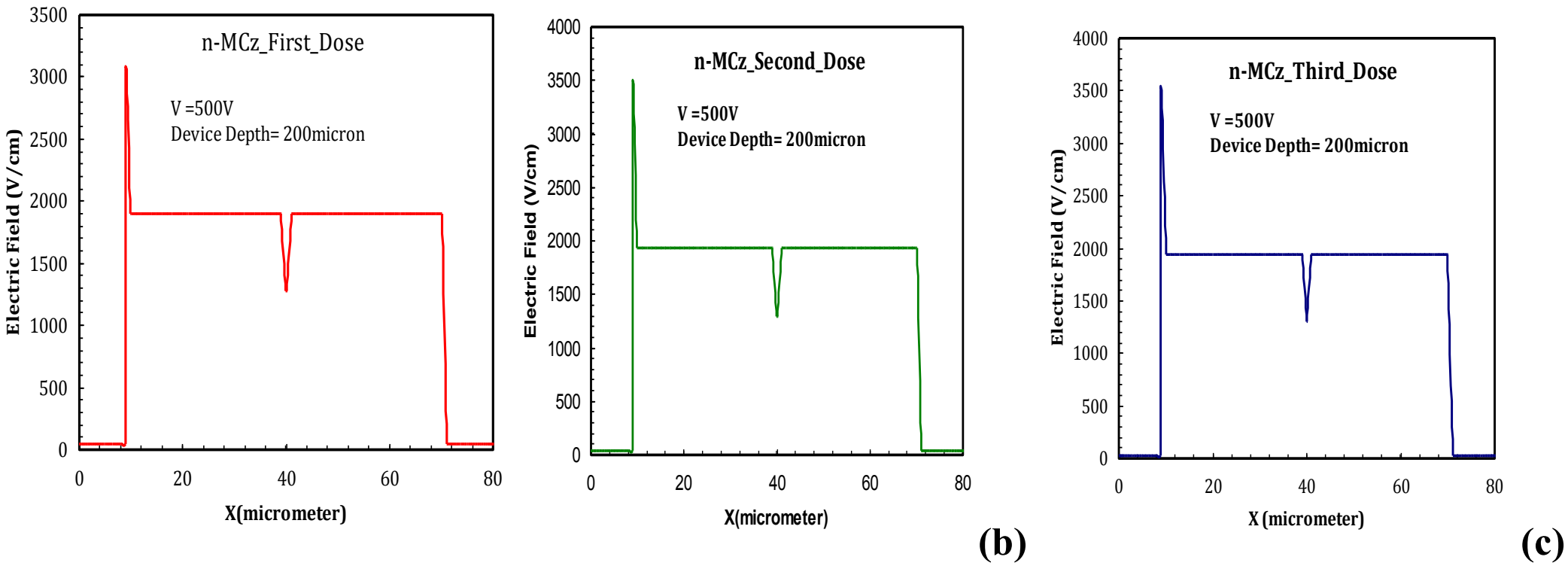

**(a)** **(b)** **(c)**

**Figure 8.** Electric Field of Mixed Irradiated n-MCz Thin Si Strip Detector for (a) 3.13 x $10^{14}$ $cm^{-2}$ (b) 4.98x$10^{14}$ $cm^{-2}$, (c)8.82x$10^{14}$ $cm^{-2}$.

It is observed form the figure 8 (a) that low E-field obtained in the base region of the detector, and E-field gutter observed in the Centre of the detector (X=40 μm, cut X=2.3 micron)**.**
It is observed from the figure 8 (b), and (c) thatE-field increases at curvature of junction and slightly increase at E-field gutter, X=40 μm

With an increasing mixed doses, E-field at curvature of junction saturates and at E-field gutter, X=40 μm tooE-field gutter can be cause of trapping of charge carrier.

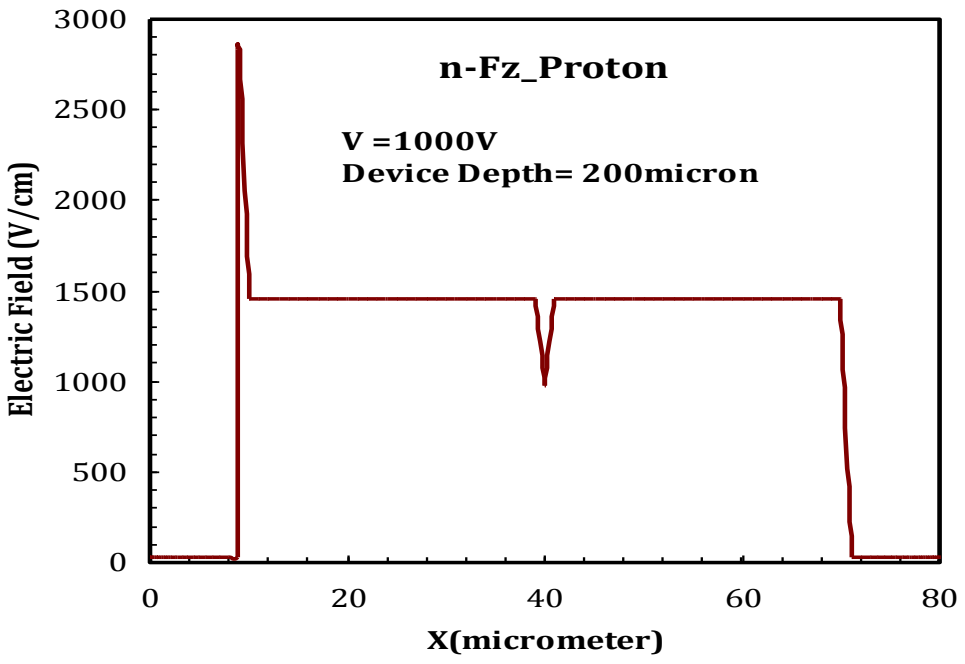

**Figure 9. Electric Field of Proton Irradiated n-Fz Thin Si Strip Detector for 2x$10^{14}$ $cm^{-2}$**

It is observed from the figure 9 less E-field at curvature of junction and in the base region of the detector, and also less E-field at E-field gutter than mixed irradiated n-MCz Si Strip detectorHigh CCE expected in thin n-MCz than n-Fz Si strip detector, traps modifying E-field in the base region of the detector.

## 5. Conclusion

In this paper we explained SRH calculation of full depletion voltage and leakage current for n-Fz/n-MCz thick/thin strip detector. Hamberg penta trap model used for proton irradiated n-Fz strip detector and four level deep trap mixed irradiated radiation damage model for n-MCz silicon strip detector. The $V_{FD}$ is the main macroscopic parameter that can determine the space charge behavior of the mixed irradiated detectors. It is observed after the SRH calulations the Full Depletion ($V_{FD}$) voltage of n-MCz thin silicon strip detector is less as compared to the thin n-Fz silicon strip detector.

. Whereas, the generation leakage current in the mixed irradiated detectors can have the uncertainty of 10-30 %, which can be due to experimental measurements.Therefore after the SRH calculation leakage current of the n-MCz thin silicon strip detector is less as compared to the thin n-Fz silicon strip detector.

It is also revealed from the TCAD simulation result 200 μm n-MCz thin Si microstrip detector consists high Charge Collection Efficiency( CCE) expected as compared to n-Fz. MGR design is also proposed for the n-MCz Si strip detector. Therefore, the radiation hard thin n-MCz Si microstrip detector for the new CMS tracker detector system can be designed and optimized for the HL-LHC experiments.